\documentclass[preprint,amsmath,amssymb,nofootinbib]{revtex4}
\usepackage{amsmath}
\usepackage{amssymb}
\usepackage{color}

\definecolor{lightgray}{rgb}{.7,.7,.7}

\definecolor{red}{rgb}{1,0,0}

\newcommand{\eps}{{\epsilon}}

\newcommand{\de}{\delta}

\newcommand{\del}{{\nabla}}
\newcommand{\mS}{{\mathcal{S}}}

\begin{document}

\title{Thermodynamics of Black Holes from Equipartition of Energy and Holography}

\author{Yu Tian}\email{ytian@gucas.ac.cn}
\affiliation{College of Physical Sciences, Graduate University of Chinese Academy of Sciences\\
Beijing 100049, China}

\author{Xiao-Ning Wu}\email{wuxn@amss.ac.cn}
\affiliation{Institute of Mathematics, Academy of Mathematics and
System Science, The Chinese Academy of Sciences, Beijing 100190,
China} \affiliation{Hua Loo-Keng Key Laboratory of Mathematics,
Chinese Academy of Sciences, Beijing 100190, China}

\date{\today}

\begin{abstract}
A gravitational potential in the relativistic case is introduced as
an alternative to Wald's potential used by Verlinde, which
reproduces the familiar entropy/area relation $S=A/4$ (in the
natural units) when Verlinde's idea is applied to the black hole
case. Upon using the equipartition rule, the correct form of the
Komar mass (energy) can also be obtained, which leads to the
Einstein equations. It is explicitly shown that our entropy formula
agrees with Verlinde's entropy variation formula in spherical cases.
The stationary space-times, especially the Kerr-Newman black hole,
are then discussed, where it is shown that the equipartition rule
involves the reduced mass, instead of the ADM mass, on the horizon
of the black hole.


\end{abstract}

\maketitle

\section{Introduction}

The discovery of black hole entropy and thermodynamics \cite{thermo}
reveals a rather general and profound relation between gravity and
thermodynamics. Later, based on the area law of entropy for all
local acceleration horizons, the Einstein equations were derived
from the first law of thermodynamics \cite{Jacobson}. Recently,
Padmanabhan reinterpreted the relation $E=2T S$ \cite{Padmanabhan}
between the Komar energy, temperature and entropy as the
equipartition rule of energy \cite{Padmanabhan2}, and Verlinde
derived the Einstein equations \cite{Verlinde} from the
equipartition rule of energy and the holographic principle
\cite{holography}. Some related works can be seen in
\cite{SG,CCO,Smolin,Makela,CM,LW,Gao,ZGZ,Culetu,Wang,Wang2,WLW,LW2,LKL,Zhao,Myung,Glikman,LWW}.

{A well-accepted fact is the entropy of black hole horizon satisfies
$S={A}/{4}$, i.e. the Bekenstein-Hawking entropy. It is interesting
to ask whether Verlinde's idea will match this well-known fact. If
the answer is positive, it will strongly support
 Verlinde's idea.} However, in Verlinde's proposal and following
works, only the change of entropy are concentrated on, while the
entropy itself is not clearly discussed. {In this paper, we try to
consider this problem.} Based on the linear superposition of the
gravitational potential in the nonrelativistic case, we find that
the entropy associated to the nonrelativistic (Laplace) horizon
naturally satisfies the familiar entropy/area relation $S=A/4$. Then
we generalize the nonrelativistic entropy formula to the
relativistic case, but it is shown that we must use a new potential,
instead of Wald's potential used by Verlinde, in order to obtain the
correct entropy/area relation $S=A/4$ on the black hole horizon.
{Because the potential plays a critical role in Verlinde's
derivation in the relativistic case, we need to check whether
Verlinde's proposal works for our new potential. First, we show that
this new potential can lead to the correct form of the Komar mass
(energy) and so the Einstein equations, as well, upon using the
equipartition rule. Furthermore, We also verify our relativistic
entropy formula by showing that it matches Verlinde's formula for
the change of entropy in two spherical cases. That evidence means
that our potential and entropy formula maybe correct.}

In the stationary case, it is obvious that there is ambiguity on the
choice of the Killing vector, which gives no contribution in the
static case. This ambiguity can be largely avoided. To be explicit,
we take the Kerr-Newman black hole as a rather general example,
where we introduce a canonical choice of the Killing vector and show
that the equipartition rule involves the reduced mass, instead of
the ADM mass, on the horizon of the black hole.

\section{Nonrelativistic Entropy}

According to Verlinde's discussion \cite{Verlinde}, if the change of
gravitational potential on the holographic screen $\mathcal{S}$
induced by the movement of a particle (with mass $m$) outside the
screen\footnote{By this term we mean that the particle is in the
already emerged space.} is $\delta\Phi$, then the corresponding
change of entropy density $\delta s$ is determined by\footnote{The
area element $dA$ is missing in eq.(4.22) of Verlinde's original
article \cite{Verlinde}.}
\begin{equation}\label{entropy_density}
  \delta s dA=-k_B \frac{\delta\Phi}{2c^2} dN,
\end{equation}
where $dN$ is the number of degrees of freedom (bits) on the
infinitesimal area element $dA$, given by
\begin{equation}
  dN=\frac{c^3}{G\hbar} dA.
\end{equation}
So we have
\begin{equation}
  \delta S=\int_\mathcal{S}\delta s dA=-\int_\mathcal{S}\frac{k_B c}{2G\hbar}\delta\Phi dA.
\end{equation}
It is easy to check that, given the Poisson equation satisfied by
the gravitational potential, a normal shift $\delta x$ of the
particle approaching the screen will induce the correct change of
entropy
\begin{equation}\label{entropy_change}
  \delta S=-2\pi k_B\frac{m c}{\hbar}\delta x,
\end{equation}
if the screen is an equipotential surface.

Since the nonrelativistic gravitational potential satisfies linear
superposition, it is reasonable to choose the entropy associated to
the screen as
\begin{equation}\label{entropy}
  S=-\int_\mathcal{S}\frac{k_B c}{2G\hbar}\Phi dA,
\end{equation}
where we have omitted a possible integration constant. If we take a
point source $M$ with the potential $\Phi=-G M/r$, then we have
\begin{equation}
  S=\int_\mathcal{S}\frac{k_B c}{2\hbar}\frac{M}{r} dA=\frac{2\pi k_B c}{\hbar} M r.
\end{equation}
On the Laplace horizon $\Phi=c^2/2$,\footnote{The nonrelativistic
horizon was taken at $\Phi=2c^2$ in Figure 4 of Verlinde's original
article \cite{Verlinde}.} the entropy becomes
\begin{equation}
  S=\frac{k_B c^3}{4G\hbar} A,
\end{equation}
which is just the familiar entropy/area relation for the
(relativistic) black hole horizon, but now in the nonrelativistic
context.

\section{Relativistic Entropy}

In the relativistic case, the explicit form of the change of entropy
density has not been discussed in \cite{Verlinde}. A natural choice
is to directly generalize eq.(\ref{entropy_density}) to the
relativistic case, i.e.
\begin{equation}
  \delta s dA=-k_B \frac{\delta\Phi}{2c^2} dN,
\end{equation}
so eq.(\ref{entropy}) becomes
\begin{equation}\label{rel_entropy}
  S=-\int_\mathcal{S}\frac{k_B c}{2G\hbar}\Phi dA,
\end{equation}
where $\Phi$ is the relativistic counterpart of the nonrelativistic potential
$\Phi$. However, if we let
\begin{equation}\label{potential}
  \Phi=\frac{c^2}{2}\ln (-\xi^2)
\end{equation}
as proposed by Verlinde, where $\xi^a$ is a time-like Killing
vector, it is easy to see that the above entropy is divergent when
the screen approaches the black hole horizon.

The above problem can be avoided by choosing an alternative
relativistic potential
\begin{equation}\label{new_potential}
  \Phi=-\frac{c^2}{2} (1+\xi^2),
\end{equation}
{where $\xi^a$ is also the time-like Killing vector normalized at
spacial infinity, i.e. $\xi^2 = -1$ at spacial infinity.} This
potential has the same asymptotic behavior as Verlinde's potential
(\ref{potential}) when the screen goes towards the spacial infinity
(near the spacial infinity, the behavior of the norm square of
$\xi^a$ is $\xi^2\to -1+\frac{2M}{r}+o(r^{-1})$ \cite{Bartnik}), but
correctly gives
\begin{equation}
  S=\frac{k_B c^3}{4G\hbar} A
\end{equation}
when the screen approaches the black hole horizon $\xi^2\to 0$.

In Verlinde's proposal, the gravitational potential plays a central
role. Using his gravitational potential and the equipartition rule
on the screen, Verlinde finds that the {quasi-local energy contained
in the screen} is just the Komar energy. This agrees with the
standard result of general relativity. Based on this result, he also
gets the Einstein equation. In order to check whether our potential
is correct, we must show that our potential (\ref{new_potential})
can also give the correct quasi-local energy (Komar mass) inside the
screen via the equipartition rule. {First, the temperature
\begin{eqnarray}
k_B T&=&\frac{\hbar}{2\pi c} N^a\nabla_a\Phi\nonumber\\
&=&-\frac{\hbar c}{2\pi} N^a\xi^b\nabla_a\xi_b\nonumber\\
&=&\frac{\hbar c}{2\pi} N^a\xi^b\nabla_b\xi_a\nonumber\\
&=&\frac{\hbar c}{2\pi}
(-\xi^2)N^a\frac{\xi^b}{|\xi|^2}\nabla_b{\xi_a}\nonumber\\
&=&\frac{\hbar c}{2\pi}
(-\xi^2)N^a\frac{\xi^b}{|\xi|}\nabla_b\frac{\xi_a}{|\xi|}-\frac{\hbar}{2\pi
c}
(-\xi^2)(N\cdot\xi)\frac{\xi^b}{|\xi|}\nabla_b\frac{1}{|\xi|}\nonumber\\
&=&\frac{\hbar c}{2\pi}
(-\xi^2)N^a\frac{\xi^b}{|\xi|}\nabla_b\frac{\xi_a}{|\xi|}-\frac{\hbar}{2\pi
c}
(-\xi^2)(N\cdot\xi)\frac{\xi^b}{|\xi|}\frac{1}{2|\xi|^3}\nabla_b(-\xi_a\xi^a)\nonumber\\
&=&\frac{\hbar c}{2\pi}
(-\xi^2)N^a\frac{\xi^b}{|\xi|}\nabla_b\frac{\xi_a}{|\xi|}-\frac{\hbar}{2\pi
c}
(\xi^2)(N\cdot\xi)\frac{\xi^b\xi^a}{|\xi|^4}\nabla_b\xi_a\nonumber\\
&=&\frac{\hbar c}{2\pi}
(-\xi^2)N^a\frac{\xi^b}{|\xi|}\nabla_b\frac{\xi_a}{|\xi|}-\frac{\hbar}{2\pi
c}
(\xi^2)(N\cdot\xi)\frac{\xi^b\xi^a}{|\xi|^4}\nabla_{[b}\xi_{a]}\nonumber\\
&=&\frac{\hbar}{2\pi c}(-\xi^2) N^b a_b\label{temperature}
\end{eqnarray}
is defined following Verlinde's proposal (but using our potential),
where $N^b$ is an outward unit vector normal to the screen, and
\begin{eqnarray}
  a^b&=&c^2 u^a\del_a u^b=c^2\frac{\xi^a}{|\xi|}\nabla_a\frac{\xi^b}{|\xi|}
\end{eqnarray}
the proper acceleration.} We interpret this temperature as measured
with respect to the reference point at spacial infinity, since the
potential $\Phi$ is defined with respect to that reference point.
But when employing the equipartition rule at the screen, we must use
the local temperature $T_\mathcal{S}$ just measured there, which is
determined by the well-know Tolman relation
\begin{equation}
  T=\sqrt{-\xi^2} T_\mathcal{S}.
\end{equation}
Then, the (generalized) equipartition rule reads
\begin{eqnarray}
  E&=&\frac 1 2\int_\mathcal{S} k_B T_\mathcal{S} dN\nonumber\\
  &=&\frac{\hbar}{4\pi c}\int_\mathcal{S}\sqrt{-\xi^2} N^b a_b
  dN\nonumber\\
  &=&\frac{c^2}{4\pi G}\int_\mathcal{S}\sqrt{-\xi^2} N^b a_b
  dA\nonumber\\
  &=&\frac{c^2}{4\pi G}\int_\mathcal{S} {\bf
  {}^2\eps}N^a\frac{\xi^b}{|\xi|}\nabla_b{\xi_a}\nonumber\\
  &=&\frac{c^2}{4\pi G}\int_\mathcal{S} {\bf {}^2\eps}N^au^b\nabla_{[b}\xi_{a]}\nonumber\\
  &=&\frac{c^2}{8\pi G}\int_\mathcal{S} *d\xi=M_{\rm Komar}(\mathcal{S},\xi),\label{energy}
\end{eqnarray}
so it is clear that the energy got from holographic principle is
indeed the Komar energy inside the screen. Then the strategy given
by Verlinde, from the expression of the Komar mass (energy) to the
Einstein equations, can be similarly followed. {So we have shown
that Verlinde's proposal also works for our new potential.}

In \cite{Verlinde}, the relativistic generalization of
eq.(\ref{entropy_change}) is taken to be
\begin{eqnarray}\label{rel_entropy_change}
\del_aS=-2\pi k_B\frac{m c}{\hbar} N_a.
\end{eqnarray}
Now we want to check the above relation from our assumptions
(\ref{rel_entropy}) and (\ref{new_potential}) in some simple
settings.

First, let us consider a Schwarzschild black hole surrounded by a
spherical thin shell. The mass of black hole is $M_i$, the (proper)
mass of the thin shell is $m$, and the radius of the shell is $R_m$.
The mass $m$ distributes uniformly on the shell. The screen $\mS$
can be located between the horizon of the black hole and the thin
shell, with radius $R$, i.e. $2M_i<R<R_m$. The shell is formed by
dusts. In fact, each particle of the shell is held by some extra
force which keeps the shell staying at its position stationarily.
Such a configuration can be realized by the following way: we splice
a Schwarzschild solution with mass $M_i$ into another Schwarzschild
solution with mass $M_o$, where $\Delta M=M_o-M_i$ can be determined
later. The boundary between these two solutions is at $r=R_m$, i.e.
the thin shell.

Because of the Birkhoff's theorem, we know that the metric should be
in natural units
\begin{eqnarray}
ds^2=\left\{\begin{array}{cc}
-(1-\frac{2M_o}{r})dt^{2}+(1-\frac{2M_o}{r})^{-1} dr^{2}+r^{2}d\Omega^{2}, & r>R_{m};\\
-C(1-\frac{2M_i}{r})dt^2+(1-\frac{2M_i}{r})^{-1}
{dr^2}+r^{2}d\Omega^{2}, & R_{m}>r>R.\end{array}\right.
\end{eqnarray}
Here $C$ is some constant. Based on the standard requirements of
constructing thin shell solutions \cite{shell}, the tangent
components of the metric on the thin shell should be continuous, so
we obtain the metric besides the thin shell as
\begin{eqnarray}
ds^2=\left\{\begin{array}{cc}
-(1-\frac{2M_o}{r})dt^{2}+(1-\frac{2M_o}{r})^{-1} dr^{2}+r^{2}d\Omega^{2}, & r>R_{m};\\
-\frac{R_{m}-2M_o}{R_{m}-2M_i}(1-\frac{2M_i}{r})dt^2+(1-\frac{2M_i}{r})^{-1}
{dr^2}+r^{2}d\Omega^{2}, &
R_{m}>r>R.\end{array}\right.\label{metric}
\end{eqnarray}
Like an ordinary thermodynamic process, let the thin shell collapse
towards the screen quasi-statically. The radius of the shell is
$R_{mo}$ at the beginning of this process and is $R_{mi}$ at the
end. Because the process is quasi-static, each middle state in the
process can be described by the metric (\ref{metric}) with different
$R_{m}$, where $R_{mi}<R_{m}<R_{mo}$. If we consider an
infinitesimal quasi-static collapsing process ($R_{mo}-R_{mi}\ll
R_{mi}$), the change of the entropy at the screen is
\begin{eqnarray}
  \de S&=&-\int_\mathcal{S}\frac{k_B}{2}\de\Phi dA\nonumber\\
  &=&-\int_{\mS}\frac{k_B}{2}\frac{d\Phi}{dR_m}\de
  R_m dA\nonumber\\
  &=&-\frac{k_B}{2} 4\pi
  R^2\frac{\Delta M}{(R_{m}-2M_i)^{2}}(1-\frac{2M_i}{R})\de R_m.
\end{eqnarray}

The energy-momentum tensor of the shell can be evaluated, by
substituting the metric (\ref{metric}) into the Einstein equations,
as
\begin{eqnarray}
T_{ab}=\frac{\Delta M}{4\pi r^2}\de(r-R_m) u_a u_b.
\end{eqnarray}
Comparing the above equation with the standard expression
\begin{eqnarray}
T_{ab}=\frac{m}{4\pi r^2\sqrt{g_{rr}}}\de(r-R_m)u_au_b
\end{eqnarray}
in terms of the mass $m$, we see that for $\Delta {M}\ll M_i$
\begin{equation}
  \Delta {M}=(1-\frac{2M_i}{R_m})^{1/2} m.
\end{equation}
Thus, the change of entropy is
\begin{eqnarray}
\de S&=&-2\pi k_B\Delta {M}\frac{R^2}{R_m^2}
(1-\frac{2M_i}{R_m})^{-2} (1-\frac{2M_i}{R}) \de
R_m\nonumber\\
&=&-2\pi k_B m\frac{R^2}{R_m^2} (1-\frac{2M_i}{R_m})^{-1}
(1-\frac{2M_i}{R})\de l,
\end{eqnarray}
where $l$ is the proper length in the normal direction of the shell.
When the thin shell is very close to the screen ($R_m\to R$), we
just obtain eq.(\ref{rel_entropy_change}).

Second, we consider another example, a charged thin shell surrounds
a Schwarzschild black hole. In this case, the gravity is balanced by
the electric force. Such solution has been studied very carefully
\cite{shell2}. Because of the Birkhoff's theorem, the metric is
\begin{eqnarray}
ds^2&=&\left\{\begin{array}{cc}-(1-\frac{2M_o}{r}+\frac{q^2}{r^2})dt^2+(1-\frac{2M_o}{r}+\frac{q^2}{r^2})^{-1}dr^2+r^2d\Omega,&r>R_m;\\
-C(1-\frac{2M_i}{r})dt^2+(1-\frac{2M_i}{r})^{-1}dr^2+r^2d\Omega,&R<r<R_m;\end{array} \right.\nonumber\\
C&=&\frac{1-\frac{2M_o}{R_m}+\frac{q^2}{R^2_m}}{1-\frac{2M_i}{R_m}}.
\end{eqnarray}
The parameters of the system satisfy \cite{shell2}
\begin{eqnarray}
m=R_m\left(\sqrt{1-\frac{2M_i}{R_m}}-\sqrt{1-\frac{2M_o}{R_m}+\frac{q^2}{R^2_m}}\
\right).\label{Mm}
\end{eqnarray}

If this shell experiences an infinitesimal quasi-static process, the
change of entropy on the screen is
\begin{eqnarray}
\de S&=&-\int_{\mS}\frac{k_B}{2}\de\Phi dA\nonumber\\
&=&-\int_{\mS}\frac{k_B}{2}\frac{1}{2}\frac{\partial C}{\partial
R_m}(1-\frac{2M_i}{r})\de R_mdA\nonumber\\
&=&-\frac{k_B}{2}4\pi R^2\frac{1}{2}\frac{\partial C}{\partial
R_m}(1-\frac{2M_i}{R})\de R_m\nonumber\\
&=&-\frac{k_B}{2}4\pi R^2\frac{1}{2}(1-\frac{2M_i}{R})\frac{(\frac{2M_o}{R^2_m}-\frac{2q^2}{R^3_m})(1-\frac{2M_i}{R_m})-(1-\frac{2M_o}{R_m}+\frac{q^2}{R^2_m})\frac{2M_i}{R^2_m}}{(1-\frac{2M_i}{R_m})^2}\de R_m\nonumber\\
&=&-\frac{k_B}{2}4\pi
R^2\frac{1}{2}(1-\frac{2M_i}{R})(1-\frac{2M_i}{R_m})^{-2}\frac{2}{R^2_m}(\Delta
M-\frac{q^2}{R_m}+\frac{M_i}{R_m}\frac{q^2}{R_m})\de R_m\nonumber\\
&=&-2\pi k_B
(1-\frac{2M_i}{R})(1-\frac{2M_i}{R_m})^{-2}\frac{R^2}{R^2_m}\left[(\Delta
M-\frac{q^2}{2R_m})-(1-\frac{2M_i}{R_m})\frac{q^2}{2R_m}\right]\de
R_m.\label{ec2}
\end{eqnarray}
In our case, the mass and charge of the shell is very small, so we
have from eq.(\ref{Mm})
\begin{eqnarray}
m=\frac{1}{\sqrt{1-\frac{2M_i}{R_m}}}(\Delta M-\frac{q^2}{2R_m}).
\end{eqnarray}
Substituting the above result into eq.(\ref{ec2}), the change of
entropy on the screen is
\begin{eqnarray}
\de S=-2\pi k_B\frac{R^2-{2M_i}{R}}{R^2_m-{2M_i}{R_m}}(m
-\sqrt{1-\frac{2M_i}{R_m}}\frac{q^2}{2R_m})\de l.
\end{eqnarray}
When the shell is very close to the screen ($R_m\to R$), the change
of entropy becomes
\begin{eqnarray}
\de S=-2\pi k_B (m-\sqrt{1-\frac{2M_i}{R}}\frac{q^2}{2R})\de l.
\end{eqnarray}
Comparing this result with eq.(\ref{rel_entropy_change}), the second
part in the bracket can be regarded as the contribution of the
electric field of the shell, which is from the electric
self-interaction.

\section{Stationary Space-Times and Black Holes}

In \cite{Verlinde}, Verlinde only considers the static case, but in
fact, similar considerations can be applied to the stationary case,
as can be seen from the previous section. As we have emphasized, the
potential plays a central role in Verlinde's proposal, which depends
on the Killing vector. In fact, Verlinde's derivation of the Komar
integral is valid for any Killing vector $\xi^a$. If there exist
more than one Killing vector, there will be ambiguity on the choice
of the Killing vector $\xi^a$. In the static case, such ambiguity
gives no contribution to the value of the Komar integral because of
the static condition. But in the stationary case, things will be a
little complex. The famous ``No-Hair" conjecture \cite{nohair} tells
us that the general non-extremal electric-vacuum stationary
space-time should be the Kerr-Newman space-time. So we focus on the
Kerr-Newman black hole. Because there is an additional Killing
vector $\partial_{\phi}$ besides $\partial_t$, there is ambiguity in
choosing the Killing vector $\xi^a$, which should be time-like at
least in a neighborhood of the screen.

Under the Boyer-Lindquist coordinates, the standard (1+3)-decomposed
form of the Kerr-Newman metric is
\begin{equation}
  ds^{2}=-\frac{\rho^{2}\Delta}{Q}dt^{2}+\frac{\rho^{2}}{\Delta}dr^{2}+\rho^{2}d\theta^{2}+\frac{Q}{\rho^{2}}\sin^{2}\theta(d\phi-a\frac{2Mr-q^{2}}{Q}dt)^{2},
\end{equation}
where $\rho^{2}=r^{2}+a^{2}\cos^{2}\theta$ with $a=J/M$ the angular
momentum per unit (ADM) mass, $\Delta=r^{2}+a^{2}-2Mr+q^{2}$ with
$q$ the electric charge, and $Q=(r^{2}+a^{2})^{2}-\Delta
a^{2}\sin^{2}\theta$. Up to a constant overall factor, the most
general form of the Killing vector $\xi$ is
$\partial_t+\omega\partial_\phi$, with $\omega$ some arbitrary
``angular velocity". The horizon is at $\Delta=0$, i.e.
\begin{equation}
  r_\pm=M\pm\sqrt{M^2-a^2-q^2}.
\end{equation}

But we see that in this case, when concerning fixed
$(r_{\mathcal{S}},\theta_{\mathcal{S}})$ outside the outer horizon,
there is a natural choice of the Killing vector $\xi$ as
\begin{equation}\label{canonical}
  \xi=\partial_{t}+a\frac{2Mr_{\mathcal{S}}-q^{2}}{Q_{\mathcal{S}}}\partial_{\phi},
\end{equation}
which is automatically time-like at the given
$(r_{\mathcal{S}},\theta_{\mathcal{S}})$ and whose norm square is
just the square of the lapse function
\begin{equation}
  f^2=\frac{\rho^{2}\Delta}{Q}.
\end{equation}
We call this $\xi$ the canonical Killing vector with respect to
$(r_{\mathcal{S}},\theta_{\mathcal{S}})$. The corresponding
acceleration is
\begin{eqnarray}
  a^{\mu} & = & u^{\nu}\nabla_{\nu}u^{\mu}=f^{-2}\xi^{\nu}\nabla_{\nu}\xi^{\mu}\nonumber\\
  & = & \frac{1}{2}f^{-2}g^{\mu\nu}\partial_{\nu}f^{2},
\end{eqnarray}
so we have
\begin{eqnarray}
  a^{r} & = & \frac{1}{2}f^{-2}g^{rr}\partial_{r}\frac{\rho^{2}\Delta}{Q}\nonumber\\
  &=&f^{-2}g^{rr}\left(\frac{r\Delta+\rho^{2}(r-M)}{Q}-\frac{\rho^{2}\Delta[2r(r^{2}+a^{2})-(r-M)a^{2}\sin^{2}\theta]}{Q^{2}}\right),
\end{eqnarray}
\begin{eqnarray}
  a^{\theta} & = & \frac{1}{2}f^{-2}g^{\theta\theta}\partial_{\theta}\frac{\rho^{2}\Delta}{Q}=f^{-2}g^{\theta\theta}\left(\frac{-a^{2}\sin\theta\cos\theta\Delta}{Q}-\frac{\rho^{2}\Delta(-\Delta a^{2}\sin\theta\cos\theta)}{Q^{2}}\right)\nonumber\\
  & = & -f^{-2}g^{\theta\theta}\Delta
  a^{2}\sin\theta\cos\theta\frac{(2Mr-q^{2})(r^{2}+a^{2})}{Q^{2}},
\end{eqnarray}
\begin{equation}
  a^\phi=0.
\end{equation}

In order to calculate the Komar energy (\ref{energy}), we must
assign a screen $\mathcal{S}$. The key point here is that the
Killing vector $\xi$ is globally defined, so it must be the same
everywhere on the screen. Thus we see that a screen adapted to the
canonical Killing vector (\ref{canonical}) is which satisfies
\begin{equation}
  b=\frac{2Mr-q^{2}}{Q}=\mathrm{const.}
\end{equation}
When $r$ ranges from $r_+$ to $\infty$, $b$ ranges from
$(r_+^2+a^2)^{-1}$ to 0. Then the Komar energy (\ref{energy}) can be
expressed as a function of $b$, but the explicit form is rather
complicated and is not necessary to present here, so we only discuss
two important cases: the infinity and the horizon. In both these
cases, the screens are spherical ($N^b\partial_b\sim\partial_r$) and
the corresponding expressions simplify drastically.
\begin{description}
  \item[Infinity] When $\mathcal{S}\to S_{\infty}$, the leading-order behavior of $a^r$
  is $a^r\approx M/r^2$, so eq.(\ref{energy}) gives $E=M$, the
  ADM mass of the black hole, as expected.
  \item[Horizon] When $\mathcal{S}\to S_H$, we have $Q\to (r_+^2+a^2)^2$ and
  \begin{equation}
    \sqrt{-\xi^2} N^b a_b\to\frac{r_+-M}{r_+^{2}+a^{2}},
  \end{equation}
  so eq.(\ref{energy}) gives $E=M_0=\sqrt{M^{2}-a^{2}-q^{2}}$, the
  reduced mass of the black hole. Note that when tending to the
  horizon, the canonical Killing vector (\ref{canonical}) is the only possible
  choice to be a time-like one. Thus we see clearly that when
  applying the equipartition rule to the black hole horizon, it is
  the reduced mass $M_0$ that takes the place of the ADM mass $M$.
\end{description}

Finally, we mention briefly two special cases where there are
simple, explicit forms of the Komar energy for generic $b$.
\begin{itemize}
  \item The $a=0$ case,
  i.e. the Reissner-Nordstrom black hole. In this case, the canonical
  Killing vector (\ref{canonical}) just becomes $\partial_t$ and there
  is no restriction on the shape of the screen, so the well-known
  result
  \begin{equation}
    E=M-\frac{q^{2}}{r}
  \end{equation}
  is recovered for general spherical screens, which tends to the
  reduced mass $M_0=\sqrt{M^2-q^2}$ when $r\to r_+=M+M_0$.
  \item The $q=0$ case, i.e. the Kerr black hole. Since the Komar
  energy (\ref{energy}) is linear in $\xi$, we have from
  eq.(\ref{canonical})
  \begin{equation}
    E=M+a b M_\phi,
  \end{equation}
  where $M$ and $M_\phi$ are the Komar integrals corresponding to
  $\partial_t$ and $\partial_\phi$, respectively. In fact, $M$ and
  $M_\phi=-2a M$ are independent of the screen in this vacuum case,
  so we obtain the (canonical) Komar energy
  \begin{equation}
    E=M-2a^2 b M.
  \end{equation}
  It is then easy to check that $E$ tends to the
  reduced mass $M_0=\sqrt{M^2-a^2}$ when $b\to (r_+^2+a^2)^{-1}$.
\end{itemize}

\section{Concluding Remarks}

In this paper, we introduce a new gravitational potential, instead
of Wald's potential used by Verlinde, in the relativistic case to
obtain the correct entropy/area relation $S=A/4$ on the black hole
horizon. Upon using the equipartition rule of energy, we also obtain
the correct form of the Komar mass (energy), which leads to the
Einstein equations. We then discuss the stationary space-times,
especially the Kerr-Newman black hole, where it is shown that a
canonical choice of the Killing vector can be defined, and the
equipartition rule involves the reduced mass, instead of the ADM
mass, on the horizon of the black hole.

In \cite{Verlinde}, Verlinde gives the formula of the change of
screen entropy caused by movement of a massive particle near the
screen. In order to check our relativistic entropy formula, we
consider two simple examples and find that our entropy formula
agrees with Verlinde's formula in both cases. However, the general
case is much more difficult, due to the high nonlinearity of the
Einstein equations. How to prove that relation in the general case
is still an open problem.

The recovery of the correct entropy/area relation in the
relativistic case is certainly helpful to clarify some subtle points
in this case in Verlinde's discussion, and to further investigate
the microscopic, or statistical, meaning of the gravitational
thermodynamics, which is left for future works.

\begin{acknowledgments}
We thank Dr. X.-D. Zhang, H.-B. Zhang, Profs. C.-G. Huang, S.-J.
Gao, M. Li, D.-S. Yang and Y.-S. Piao for helpful discussions. This
work is partly supported by the National Natural Science Foundation
of China (Grant Nos. 10705048 and 10731080) and the President Fund
of GUCAS.
\end{acknowledgments}

\end{document}